\documentclass[usenatbib]{mn2e}
\usepackage{epsfig}
\usepackage{amsmath}
\usepackage{wasysym}
\usepackage{natbib}
\usepackage{graphicx}
\usepackage{epstopdf}

\title[Fluctuations in High-Redshift Radiation Backgrounds]{Fluctuations in the High-Redshift Lyman-Werner and Lyman-$\alpha$ Radiation Backgrounds}

\author[Holzbauer \& Furlanetto]{Lauren N. Holzbauer\thanks{Email: holzbauer@astro.ucla.edu} and Steven R. Furlanetto \\ 
Department of Physics \& Astronomy, University of California Los Angeles; Los Angeles, CA 90095, USA} 

\begin{document}

\maketitle

\begin{abstract}
We use a new method to model fluctuations of the Lyman-Werner (LW) and Lyman-$\alpha$ radiation backgrounds at high redshift. At these early epochs the backgrounds are symptoms of a universe newly lit with its first stars. LW photons (11.5-13.6 eV) are of particular interest because they dissociate molecular hydrogen, the primary coolant in the first minihalos. By using a variation of the halo model, we efficiently generate power spectra for any choice of radiation background. We find that the LW power spectrum typically traces the matter power spectrum at large scales but turns over at the scale corresponding to the effective `horizon' of LW photons ($\sim 100$~comoving Mpc), unless the sources are extremely rare. The series of horizons that characterize the Lyman-$\alpha$ flux profile shape the fluctuations of that background in a similar fashion, though those imprints are washed out once one considers fluctuations in the brightness temperature of the 21-cm signal. The Lyman-$\alpha$ background strongly affects the redshifted 21-cm signal at just about the time the LW background begins to dissociate H$_2$, so measuring that background's properties will reveal important information about the transition from early Population III stars to more normal stars. Around this time we find that fluctuations in the LW background are weak; the fractional standard deviation is less than $\sim 0.5$ on scales $\apprge 10$ cMpc, only rising to be of order unity on scales $\apprle 1$ cMpc. This should not lead to substantial spatial fluctuations in H$_2$ content, except at the earliest times. Even then, most halos form far from other sources, so the transition from star formation in low-mass to high-mass halos is rather homogeneous across the universe.
\end{abstract}


\section{Introduction}
An important aspect of the cold dark matter (CDM) universe is that density fluctuations exist on small scales. These small scale perturbations are superimposed on larger scale perturbations; the density reaches its highest value over the smallest region. Consequently, structure forms via hierarchical buildup. An initially smooth density distribution eventually morphs into a web of sheets and filaments. It is the overdense junctions of these filaments that we call dark matter halos. Further structure development takes place inside these halos, commencing with the first (Population III) stars.

Population III (Pop III) stars illuminated our dark universe in its cold youth and from them developed the complex environment we live in today. According to hierarchical structure formation, these stars formed out of metal-free H/He gas contained in minihalos at redshifts $z  \sim  20$ - $30$~\citep{couchmanrees1986}. The minihalos, with masses around $10^6 M_{\odot}$, have virial temperatures less than $10^4$ K - below the threshold for atomic hydrogen cooling~\citep{ohhaiman2002}. Consequently, the halos have to rely primarily on H$_2$ for cooling~\citep{htl96, tegmark97, abel2002, bromm2002}. This cooling takes place via collisional excitation (mainly between H$_2$ molecules and energetic H atoms) and subsequent radiative decay of the rotational transitions of H$_2$. 

In the classical view of Pop III star formation, molecular cooling produces a single, massive star from cold gas that becomes trapped in the dark matter potential well of one minihalo~\citep{bromm2009}. A dense core, or protostar, gradually emerges and grows into a massive star by accreting the surrounding gas. These first stars could theoretically grow to be several hundred solar masses~\citep{brommloeb2004}; most were probably $\sim 100 M_\odot$~\citep{brommlarson2004}. But what would happen if the infalling gas became fragmented? Several studies show that a primordial protostellar cloud will most likely not violently fragment enough for a secondary clump to compete with the parent clump and form a second star~\citep{abel2002, yoshida2008}. However very recent simulations suggest that if a gas cloud surrounding a protostellar core has an initial degree of angular momentum, it could collapse into a dense disk, cool, and fragment, resulting in a binary or even multiple Pop III star system~\citep{stacy2010b, turk2009} consisting of two or more lighter stars as opposed to a single, massive star. These new studies could indicate that the formation of the first stars could be more complicated and varied than previously believed.

In these primordial star cookers, the ability to form new stars is dependent on the abundance of H$_2$. However, as the population of these luminous stars grows, the sites of star formation are increasingly irradiated by soft UV photons (the Lyman-Werner, or LW, bands: 11.2-13.6 eV) from existing stars. This LW radiation can photodissociate the H$_2$ molecules in the gas through the two-step Solomon process~\citep{field66, stecher67},
\begin{equation}
	{\rm H}_2 + \gamma \rightarrow {\rm H}^{*}_{2} \rightarrow 2{\rm H},
\end{equation}
in which an H$_2$ molecule hit by a LW photon bumps it up to an excited electronic state, H$^{*}_{2}$. A fraction of decays from this excited state end up in the vibrational continuum of the ground state, dissociating the molecule. If the LW background becomes strong enough, it can prevent further collapse and consequently stall the further formation of primordial ionizing sources by terminating the minihalos' primary cooling supply~\citep{hrl97}. As a result, the only halos able to cool (via atomic line cooling) and form new stars are those with $T_{\rm vir} \ga 10^4$ K, or masses above $\sim 10^8 M_\odot [(1+z_{\rm vir})/10]^{-3/2}$. Minihalos, with temperatures below the threshold for atomic cooling, will not be able to collapse past virialization without a sufficient supply of H$_2$.

Although these larger halos may still form metal-free stars, the thermodynamics of the cooling process is sufficiently different that we expect the resulting stars to differ substantially (especially in their characteristic mass).  The two populations are sometimes described as Population III.1 and Population III.2 to emphasize this: both may be `primordial,' but they have very different properties regulated by the LW (and other) radiation backgrounds (see, e.g., \citealt{bromm2009}).

Most previous calculations of the LW background used a homogeneous approximation, in which they assumed a uniform distribution of sources~\citep{haiman00, ricotti2002, yoshida2003}. But the highly clustered, discrete sources responsible for the background radiation do not generate a uniform background. If these fluctuations are large enough, the transition from H$_2$ cooling to atomic line cooling would be very patchy, potentially allowing exotic star formation to persist for long periods even after the mean background reaches the threshold value for H$_2$ suppression. \citet{dijkstra2008} were the first to consider the inhomogeneous LW background, but only in the context of close halo pairs (using a Monte Carlo model) and only when the background was already well above threshold. ~\cite{ahn} were the first to consider the inhomogeneous background using a large-scale radiative transfer simulation of reionization.

These photons have other observable effects as well; most importantly, as they redshift into the Lyman-$\alpha$ transition they couple the excitation temperature of the 21-cm transition of hydrogen to the gas kinetic temperature via a radiative pumping mechanism known as the Wouthuysen-Field effect \citep{wouthuysen1952, field1959a}. This renders the 21-cm signal visible in emission or absorption.  ~\cite{barkloeb2005} showed that the fluctuations in this young Lyman-$\alpha$ background produced strong fluctuations in the 21-cm signal. Conversely, observing these fluctuations can reveal a wealth of information as to the properties of these first luminous sources. We will see that these photons begin to affect the 21-cm background at roughly the same background intensity at which they suppress H$_2$ cooling.  Thus redshifted 21-cm measurements offer an excellent chance to study the transition from Population III.1 to III.2 stars as well as the inhomogeneities in the ultraviolet radiation field during the `cosmic dawn.'

In this paper, we present a new method with which to efficiently calculate the power spectrum of an arbitrary radiation field for any desired redshift and range in scale (in this paper we focus on the LW and Lyman-$\alpha$ backgrounds specifically; see \citealt{mesfurl2009} for an earlier application specific to the ionizing background). Using the halo model to determine the spatial distribution of halos, we can build up the radiation background by superimposing a flux profile specific to that particular background on each halo. This profile effectively replaces the mass density profile traditionally used in the halo model to calculate fluctuations in the density field. We aim to study the importance of fluctuations in these backgrounds and complement the radiative transfer simulation of~\cite{ahn} with our simple, analytic model. Our method also takes a very different approach to calculating 21-cm fluctuations (due to perturbations in the Lyman-$\alpha$ radiation field) compared to existing work \citep{barkloeb2005, pritchardfurl}.

In this first exploration of the radiation background, we restrict our attention to the soft-UV background from a relatively simple model of first galaxy formation.  In fact, many other physical factors contributed to the transition from Population III to Population II star formation.  The most obvious is metal enrichment, which also affects the cooling and is highly inhomogeneous (see, e.g., \citealt{furlloeb2005}).  Also, X-rays emanating from the first sources can counteract H$_2$ destruction by increasing the free electron fraction and so catalyzing its formation~\citep{mcdowell61, hrl96, haiman00}. There has been considerable debate as to which of these backgrounds is more influential. For our simple model, we will follow ~\citet{machacek2003} by assuming that the enhancement of the electron density due to the X-ray background occurs too slowly to compete with photodissociation and so neglect the X-ray background.

Recently, \citet{tseliakhovich2010} pointed out that the residual relative velocities of the baryon fluid and underlying dark matter distribution, imprinted during the recombination era by the baryons' close coupling to photons and now visible as baryon acoustic oscillations, may have important implications for star formation in these early, fragile halos.  These large-scale velocities will suppress the accretion of gas onto small dark matter halos \citep{tseliakhovich2010, tseliakhovich2010b}.  The actual implications for star formation are as yet unclear; the first simulations show modest effects on the reionization era itself \citep{maio2010,stacy2010}, but the effect on the earlier epochs is important for the LW and Lyman-$\alpha$ backgrounds \citep{dalalpen10}.  Because the effects are as yet unclear, we will ignore these velocity corrections here, thus providing a baseline prediction for comparison with future work better incorporating them.

This paper is organized as follows: in Section \ref{method} we describe our method for calculating the power spectrum of the LW and Lyman-$\alpha$ radiation background fluctuations using the halo model. In Sections \ref{lw} and \ref{lya} we calculate the flux profiles for the LW and Lyman-$\alpha$ backgrounds, respectively, and also present our results. We summarize our results and conclude in Section \ref{summary}. We adopt a background cosmology $(\Omega_0, \Omega_\Lambda, \Omega_b, h, \sigma_8,n)=(0.26, 0.74, 0.044, 0.74, 0.8, 0.95)$ consistent with the most recent measurements \citep{komatsu2011}.

\section{Method}\label{method}

We are interested in modeling the power spectrum of fluctuations in the LW and Lyman-$\alpha$ radiation backgrounds using the halo model. Unlike earlier treatments of the LW background, we are specifically interested in its large-scale inhomogeneities, complementing the high resolution, large-scale N-body radiative transfer simulation of \citet{ahn} and the small-scale treatment of \citet{dijkstra2008}. 
Instead we will expand the model of ~\cite{mesfurl2009}, who treated the inhomogeneous hydrogen-ionizing ultraviolet background using a halo model-like prescription.

The halo model, as described in~\cite{cooraysheth}, uses properties of virialized dark matter halos to calculate the effects of non-linear gravitational clustering, assuming that all mass in the universe is compartmentalized in such halos, whose properties can be parameterized purely by their mass $m$. The three ingredients of the model are the (1) halo number density, $n(m)$, (2) spatial distribution of the halos, and (3) distribution of mass within each halo, or halo density profile, $\rho(r|m)$, (where $r$ is the distance away from the center of a halo with mass $m$). Typically, a theoretically-motivated halo mass function for (1) (the classic choice being~\citealt{pressschechter}) allows one to calculate (2). The density profile for (3) can be calibrated by numerical simulations, such as the NFW~\citep{nfwref} or \citet{m99ref} profiles. This model is very powerful in that it can efficiently determine the power and many other useful properties for any density field at an arbitrary epoch and scale. 

Since we wish to quantify the radiation background rather than the mass density field, we simply replace the `halo density profile' with the profile of the radiation field around each halo. This flux profile, $\rho_{\rm rad}(r|m)$, depends on the radiation background under consideration and will be discussed later. For a spherically symmetric profile, the normalized Fourier transform, $u(k|m)$, can be written as:
\begin{equation}\label{profreduced}
u(k|m) = \frac{\int_{0}^{r_c} dr4\pi r^2 [\sin (kr)/(kr)] \rho_{\rm rad}(r|m)}{\int_{0}^{r_c} dr4\pi r^2 \rho_{\rm rad}(r|m)},
\end{equation}
where $r_c$ is the cutoff distance at which an observer can no longer see the radiation emanating from the source. For the case of the LW radiation, for example (described more fully in section~\ref{lwprofile}), this cutoff distance, or horizon, is given by $r_{\rm LW} \sim 100$ comoving Mpc (cMpc). 

As described above, we are interested in modeling the fluctuations in a variety of radiation backgrounds (in this paper, the LW and Lyman-$\alpha$ backgrounds). We use the power spectrum, $P(k)$, or the dimensionless quantity $\Delta(k) \equiv k^3P(k)/2\pi^2$ to quantify these fluctuations. Following the halo model, we write the power as a sum of two terms: the first term, $P^{\rm 1h}(k)$, describes the case for which radiation at two points comes from the same source,\footnote{In the context of the halo model, our radiation background calculation is analogous to the dark matter density power spectrum in the halo model, \emph{not} to the (discrete) galaxy power spectrum.  Thus the `one-halo' term is very important to our results on small scales, even if each halo contains only one galaxy.} while the second term, $P^{\rm 2h}(k)$, describes the case for which the two points are illuminated by two sources:
\begin{eqnarray}
P(k) & = & P^{\rm 1h}(k) + P^{\rm 2h}(k), \mbox{where} \\
P^{\rm 1h}(k) & = & \int_{M_{\rm min}}^{\infty} dmn(m) \left(\frac{m}{\bar \rho f_{\rm coll}} \right)^2 |u(k|m)|^2 \\
P^{\rm 2h}(k) & = & \left[ \int_{M_{\rm min}}^{\infty} dmn(m) \left(\frac{m}{\bar \rho f_{\rm coll}} \right) u(k|m) b(m) \right]^2 P^{\rm lin}(k).
\label{ps_halomodel}
\end{eqnarray}
Here, $\bar \rho$ is the matter density, $f_{\rm coll}$ is the collapse fraction (fraction of mass in the universe contained in galaxies, or collapsed in a halo), $b(m)$ denotes the halo bias (describing how strongly clustered the halos are; ~\citealt{mowhite96}), and lastly $P^{\rm lin}(k)$ is the linear power spectrum. Here we have approximated the halo-halo power spectrum, for two halos with mass $m_1$ and $m_2$, as $b(m_1) b(m_2) P^{\rm lin}(k)$, which requires that the halo fluctuations remain linear on the appropriate scales (i.e., those on which $P^{\rm 2h}$ dominates).  While the density fluctuations themselves are very weak at the redshifts of interest to us, the halos are also highly biased, so nonlinear corrections will be important on sufficiently small scales. We use the~\cite{eisensteinhu} fit to the transfer function to calculate $P^{\rm lin}(k)$ and the Sheth-Tormen mass function and collapse fraction~\citep{shethtormen}.

To model the density power spectrum one must include the entire halo population, over all masses. However, we are interested in the total radiation field and so should not include the low-mass halos unable to host stars. We assume that only halos more massive than a cutoff mass, $M_{\rm min}$, host stars and so contribute to the radiation background. To motivate our choices for $M_{\rm min}$, we first consider the `filter mass,'
 $M_{\rm filter}$, the characteristic scale over which baryonic perturbations are smoothed in linear perturbation theory or the minimum mass of a halo to accrete baryons~\citep{gnedinhui1998, naozbarkana2007}, as a lower limit ($\sim 10^5 M_\odot$ in our redshift regime). In linear theory, the relative force balance between gravity and pressure can be characterized by the Jeans mass, $M_J$; the corresponding Jeans scale is the minimum scale on which a small perturbation will grow due to gravity. $M_J$ depends on the instantaneous value of the sound speed of the gas, consequently overestimating the characteristic mass scale by up to an order of magnitude~\citep{gnedin2000}. In contrast, $M_{\rm filter}$, which takes into account the full thermal history of the gas, is a more accurate mass scale. 
 
 An upper limit would be the threshold for atomic cooling, $T_{\rm vir} \sim 10^8K$~\citep{ohhaiman2002}. Since the H$_2$ fraction, $f_{\rm H_2}$, increases with halo mass ($f_{\rm H_2} \propto T_{\rm vir}^{1.5}$,~\citealt{tegmark97}) the cutoff mass certainly lies somewhere between these two limits. The classical criterion that the cooling time be smaller than the dynamical time will set the redshift-dependent transition: in the absence of a LW background, these successful minihalos probably have $f_{\rm H_2} \sim 10^{-4}$ and $M_{\rm halo} \sim 10^6 M_\odot$~\citep{htl96, tegmark97, yoshida2003}. However, rather than try to model this in detail we will employ a variety of selections for $M_{\rm min}$ in order to remain most general.

\medskip

\section{The LW Background}\label{lw}

In this section we will apply the above method to fluctuations in the LW radiation background, which determines if the sterilization of minihalos at high redshift (through the photodissociation of H$_2$) was a patchy or homogeneous transition. 

\subsection{The Flux Profile}\label{lwprofile}

\begin{figure}
\begin{center}
\resizebox{8cm}{!}{\includegraphics{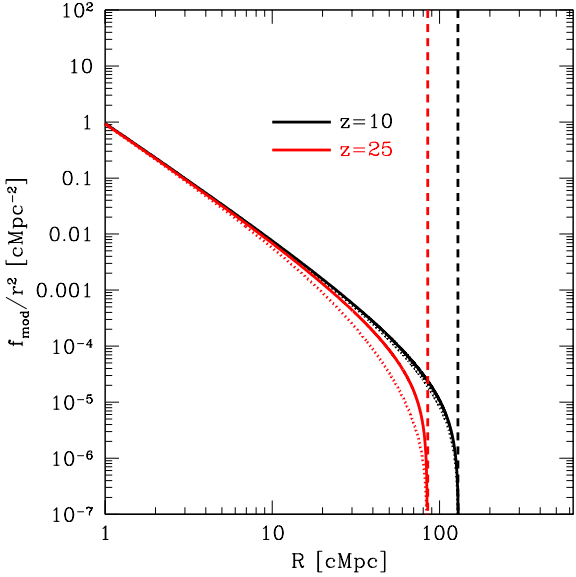}}\\
\end{center}
\caption{The normalized LW flux profile, $\rho_{\rm LW}(r|m)$, shown for $z=10, 25$ (right and left sets of curves, respectively). Each profile terminates at the horizon $r_{\rm LW}$, indicated by the vertical dashed lines. The dotted curves represent the corresponding time dependent versions of the flux profile using the timescale of typical halo growth, $t_\star$, for $M_{\rm min}=10^8 M_{\odot}$.}
\label{fmod}
\end{figure}

Our LW flux profile (shown in Figure~\ref{fmod} for $z = 10$ and $25$) for a halo with mass, $m$, located at an effective luminosity distance, $r$, from the observer is given by:
\begin{equation}
\rho_{\rm rad}(r|m) \propto m \frac{f_{\rm mod}(r)}{4\pi r^2},
\end{equation}
where we assume for simplicity that the luminosity of each halo scales with its mass. \footnote{A more complex relationship is likely, but any such relationship can be bracketed by our different choices of $M_{\rm min}$.}  Here we use the picket fence modulation factor, $f_{\rm mod}(r)$, from~\cite{ahn}. This is the fraction of LW continuum radiation emitted by a source that is received by the observer without redshifting into a hydrogen Lyman series resonance line, where it will either be absorbed or scattered. An absorbed photon will either cascade to the 2$p$ level and produce a Lyman-$\alpha$ photon or cascade to the metastable 2$s$ level and decay by two photon emission~\citep{pritchardfurl}-- either way, the resulting photon will be below the LW range. On the other hand, the scattered photon will be reabsorbed until it, too, decays into a low-frequency photon (typically after just a few scatterings). So, for a given photon at observed frequency, $\nu_{\rm obs}$, we can define a maximum redshift, $z_{{\rm max,}i}$, corresponding to the maximum distance within which photons from a source remain in the LW band without redshifting into the closest Lyman line from above (located at frequency $\nu_i$):
\begin{equation}\label{zmax}
\frac{1+z_{{\rm max,}i}}{1+z_{\rm obs}} = \frac{\nu_i}{\nu_{\rm obs}}.
\end{equation}

With each Lyman line associated with its own $z_{{\rm max,}i}$ and the spacing between them decreasing with increasing $\nu_i$, we are left with a transmission spectrum resembling a poorly fashioned picket fence, illustrated in Figure 2 from~\cite{ahn}. The modulation factor, $f_{\rm mod}$, is defined as the fraction of the LW frequency interval, 11.5-13.6eV, that lies within the pickets, or that is successfully transmitted to the observer:
\begin{equation}\label{fmodsigma}
f_{\rm mod} = 1 - \sum_j \left( \frac{h\Delta\nu_{{\rm gap,}j}}{2.1 \ {\rm eV}} \right),
\end{equation}
where $\Delta\nu_{{\rm gap},j}$ is the frequency interval between each picket in which there is no transmission. \footnote{We implicitly assume a flat photon spectrum within the LW range here, which is a reasonable approximation over this short frequency interval.} The profile terminates at the `LW horizon,' $r_{\rm LW} = 97.39 \alpha$ cMpc, the distance at which a photon redshifts across the maximum picket spacing (between the pickets corresponding to the Ly$\delta$ and Ly$\gamma$ lines). The scaling factor, $\alpha$, is defined as:
\begin{equation}\label{alpha}
\alpha = \left(\frac{h}{0.7}\right)^{-1}\left( \frac{\Omega_m}{0.27}\right)^{-1/2}\left( \frac{1+z}{21}\right)^{-1/2}.
\end{equation}

 While $f_{\rm mod}$ can be calculated numerically,~\cite{ahn} have devised a fitting formula:
\[
f_{\rm mod}(r) = 1.7\exp{\left[-(r_{\rm{cMpc}}/116.29\alpha)^{0.68}\right]} - 0.7 
\]
if $r_{\rm{cMpc}}/\alpha \le 97.39$ and zero otherwise, where $r_{\rm{cMpc}}$ is the distance to the source in cMpc. We have successfully reproduced $f_{\rm mod}$ using the method described by~\cite{ahn} and have confirmed that the fitting formula is accurate to within 2 per cent error of the true numerical values.

\subsubsection{The Light Cone}\label{timedep_lw}

There is one difficulty with the halo model as usually constructed for our problem: it does not allow the sources to evolve over time.  As usually constructed, the halo model takes the properties of each halo at a particular instant. This is not actually appropriate for our application, where the time delay from the finite speed of light implies that many sources will only be visible to a given point as they were long in the past, when their luminosity may have differed from the present value. For example, the light travel time across the LW horizon is 38.6 Myr at $z=10$ and 11.1 Myr at $z=25$. These values are a full $\sim 5\%$ of the Hubble time at those epochs, so a fraction of the visible sources would appear much dimmer than the above model would suggest. We next estimate how much we would expect the inclusion of such a time dependence to alter our results. 

We can crudely account for halo growth by attaching a damping factor to the flux profile:
\begin{equation}\label{timeprofile}
\rho_{\rm rad}(r | m) = L(m)\frac{f_{\rm mod}}{4\pi r^2} \longrightarrow L(m)\frac{f_{\rm mod}e^{-r/r_{\star}}}{4\pi r^2},
\end{equation}
where $r_\star = ct_\star (1+z)$ in cMpc and $t_\star$ corresponds to the typical timescale for halo growth, assuming that at these high redshifts the halos grow exponentially fast so that the luminosity of a halo $L(m) \propto \exp^{t/t_\star}$. The growth timescale we define as: $t_\star = a(z)t_{\rm H}$ (where $a$ is some proportionality factor that evolves over time and $t_{\rm H}$ is the Hubble time). 

To estimate $a$, consider a population of identical halos with mass $m$, at some redshift $z$, in which the halos are conserved; no new halos are created and none are destroyed. In this simple case, the collapse fraction is given by:
\begin{equation}
f_{\rm coll}(t) = \frac{n\cdot m(t)}{\bar\rho}.
\end{equation}
Note how the halo mass is now a function of time, $t$. Taking the time derivative of this expression leads to:
\begin{equation}
\frac{1}{f_{\rm coll}(t)}\frac{df_{\rm coll}}{dt} = \frac{1}{m}\frac{dm}{dt} \equiv \frac{1}{t_{\star}}.
\end{equation}
Thus for example, $a(z=10) = 0.283, 0.167$ for $M_{\rm min} = 10^6 M_\odot, 10^8 M_\odot$ respectively and $a(z=25) = 0.062, 0.034$ for the same choices of $M_{\rm min}$.

Although this simple model overestimates the growth rate of individual sources (because in reality much of the increase in collapse fraction is driven by new halos passing the relevant mass threshold), it provides a simple conservative parameterization of the effects of growth.

Adding this rapid source evolution effectively damps the source profile whenever $r_\star \la r_{LW}$ -- these effects are illustrated in Figure ~\ref{fmod}. For example, for $M_{\rm min} = 10^8 M_\odot$, $r_\star(z=10) = 401.6$ cMpc and $r_\star(z=25) = 53.7$ cMpc while $r_{\rm LW} = 129.7$ cMpc and $84.4$ cMpc for those respective redshifts. We include this crude model for the light cone effect below but also point out where it modifies our results substantially.

\subsection{The Threshold Intensity}\label{j21}

\begin{figure}
\begin{center}
\resizebox{8cm}{!}{\includegraphics{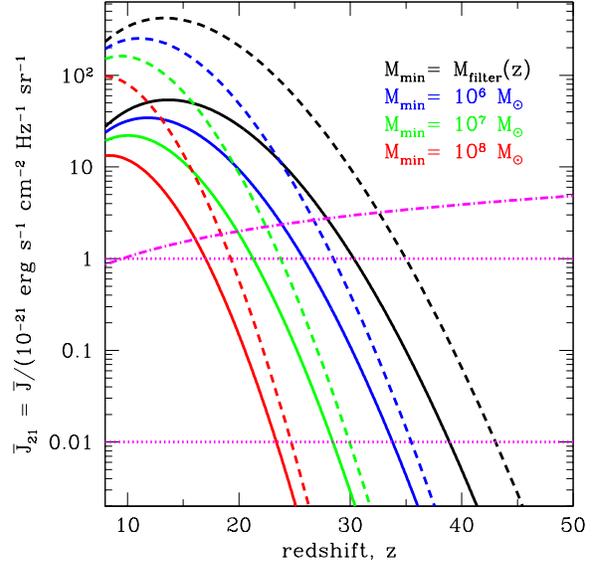}}\\
\end{center}
 \caption{The mean intensities of the LW background (solid curves), $\bar J_{\rm LW,21}$, and the Lyman-$\alpha$ background (dashed curves), $\bar J_{\alpha,21}$, for $M_{\rm min} = M_{\rm filter},\, 10^6 M_{\odot},\, 10^7 M_{\odot}$, and $10^8 M_{\odot}$ (right to left). Horizontal dotted lines indicate the upper and lower limits for the expected threshold value of $J_{\rm LW,21}$. The dot-dashed line indicates the critical intensity for the Lyman-$\alpha$ background, $J_{\alpha,21}^c$ (see \S \ref{mean21cm}).}
\label{j21_allz_allm}
\end{figure}

In order to determine when fluctuations are most important, we next compute the evolution of the LW intensity -- and hence the point at which H$_2$ cooling is suppressed -- in some simple models of structure formation.  This section is not meant to provide a detailed model of star formation, but it should provide some context for the fluctuations we will later examine. We can estimate the mean LW intensity, $\bar{J}_{\rm LW}(z)$, with the following:
\begin{equation}\label{jlw}
\bar J_{\rm LW} (z) = \frac{(1+z)^2}{4\pi} \int_{z}^{z+z_{\rm LW}} \frac{c dz'}{H(z')} \bar\epsilon (z') f_{\rm mod}(z' - z),
\end{equation}
where $f_{\rm mod}(z' - z)$ is part of the LW flux profile (described more fully in \S \ref{lwprofile}) and the mean emissivity, $\bar\epsilon(z')$, is given by:
\begin{equation}\label{meanemissivity}
\bar\epsilon (z) = f_{\star} \bar n_b^0 \frac{d}{dt}f_{\rm coll}(z) \epsilon_b,
\end{equation}
with $f_\star$ being the star forming efficiency (fraction of baryons that actually form stars), which we take to be $10\%$ as a fiducial value, and $\bar n_b^0$ is the mean baryon number density. We can approximate the spectral distribution function (defined as the number of photons per frequency $\nu$ emitted per baryon), $\epsilon_b(\nu)$, as its mean value $\epsilon_b$ over the LW range (11.2--13.6 eV) for simplicity since we are looking at such a small range in frequency. We normalize $\epsilon_b(\nu)$ to produce $4800$ photons per baryon between Lyman-$\alpha$ and the Lyman limit for very massive Population III.1 (zero-metallicity, $M \geq 100M_\odot$) stars~\citep{barkloeb2005}. Note that the light cone effect is inherent in this expression due to the redshift dependence of the mean emissivity.

Our results are summarized in Figure~\ref{j21_allz_allm}. The LW intensity is calculated in units of $J_{\rm LW,21} = J_{\rm LW}/(10^{-21}$ erg s$^{-1}$ cm$^{-2}$ Hz$^{-1}$ sr$^{-1})$. The solid curves represent the very massive Population III stars. Replacing the emissivity with that of a Population II star (with metallicity equal to $1/20$ the solar value) producing $9690$ photons per baryon in our frequency range~\citep{barkloeb2005} boosts the intensities by a factor of $\sim 2$; the background reaches threshold earlier. According to~\cite{haiman00}, background intensities of $J_{\rm LW,21} \sim 10^{-2}$ -- $1$ are needed to suppress H$_2$ cooling in all minihalos over a range of redshift from $z \sim 10 - 50$. The lower value describes H$_2$ suppression in halos near the $T_{\rm vir} < 10^{2.4}$ K limit; below these low temperatures H$_2$ cooling is inefficient even in the absence of any photodissociating background, so no stars will form. Since $f_{\rm H_2}$ increases with $T_{\rm vir}$, it will be more difficult to terminate a more massive halo's larger cooling supply. Thus, as the halo population evolves, becoming more numerous and more massive with time, the threshold intensity must increase, self-regulating star formation by shifting the minimum mass to higher values. Once halos reach $T_{\rm vir} > 10^{3.8}$ K the value of the background intensity is once again irrelevant since these large halos are able to cool via atomic line cooling and no longer rely on their fragile H$_2$ supply. In further calculations, we will take $J_{\rm LW,21} = 0.1$ as our fiducial threshold intensity.

It is evident from Figure~\ref{j21_allz_allm} that our models reach this threshold between redshifts $z \sim 15 - 35$. Of course, these models are extremely naive and ignore a host of complications (such as the evolving star formation efficiency and cooling threshold, as well as other feedback mechanisms). But they suffice to illustrate approximately when a given model reaches the H$_2$ photodissociation threshold, where the fluctuations which we will study are particularly interesting. Note that, because structure formation itself proceeds exponentially fast at high redshifts, uncertainties in the star formation parameters themselves are relatively unimportant.  In any case, $J_{\rm LW,21} \propto f_\star \epsilon_b$, so it is easy to read off the appropriate intensity for such a model.

We emphasize that the fluctuations will be most important near the threshold, because that is when the transition in cooling modes actually occurs. If fluctuations are small, the transition would occur uniformly over the entire Universe.  If not, the Universe could contain isolated, sparsely populated patches in which H$_2$ cooling remains possible. If a minihalo inhabits one of these `safe' patches, it could continue to form very massive stars via H$_2$ cooling even after the mean intensity reaches threshold. On the other hand, even well before an average IGM point reaches threshold, regions near existing sources will be well above it, and this could strongly affect the highly-clustered early sources.  We will examine this phase in \S \ref{earlyflucs}.

\subsection{Power Spectrum}\label{powerspec}

\begin{figure}
\begin{center}
\resizebox{8cm}{!}{\includegraphics{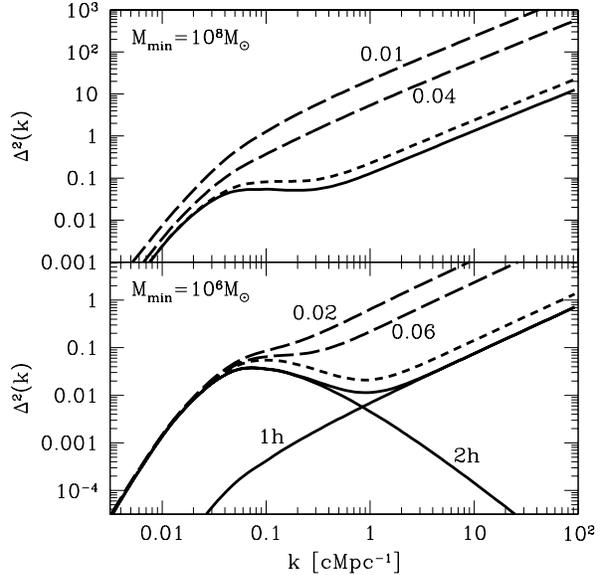}}\\
\end{center}
\caption{Power spectrum of the LW background, $\Delta^2 (k)$ for scenarios normalized to reach $J_{\rm LW,21} = 0.1$. Scenarios include: $M_{\rm min} = 10^6 M_\odot$, $z=30.15$ (bottom panel) and $M_{\rm min} = 10^8 M_\odot$, $z=20.5$ (top panel). The light cone versions for both scenarios are also shown (short-dashed curves) while the original versions are the solid curves. For the $M_{\rm min} = 10^6 M_\odot$ curve we have shown not only the total spectrum but also the one-halo (labeled `1h') and two-halo (labeled `2h') contributions for visual aid (see equation~\ref{ps_halomodel}). The two, topmost, long-dashed curves represent the light cone scenarios with the duty cycle taken into account. The $M_{\rm min} = 10^6 M_\odot$ model includes two different values for this: $f_{\rm duty} = 0.02$ (topmost curve, labeled $0.02$) and $f_{\rm duty} = 0.06$ (second curve, labeled $0.06$). The $M_{\rm min} = 10^8 M_\odot$ case also includes two curves for $f_{\rm duty} = 0.01$ and $0.04$.}
\label{delta2lw_z10}
\end{figure}

\begin{figure}
\begin{center}
\resizebox{8cm}{!}{\includegraphics{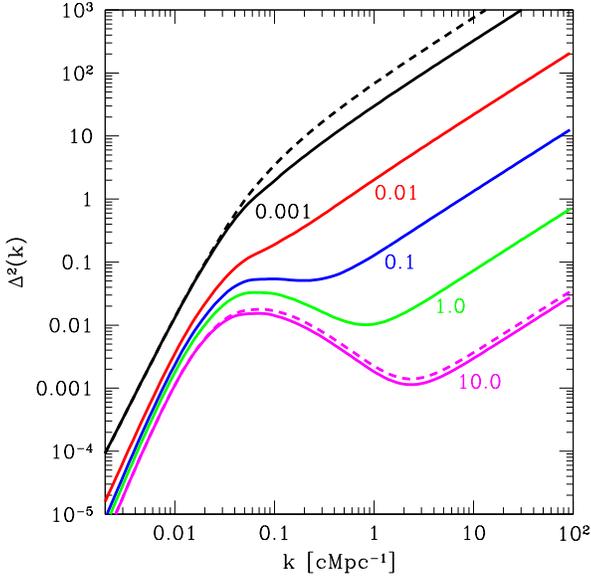}}\\
\end{center}
\caption{Power spectrum of the LW background, $\Delta^2 (k)$, for our $M_{\rm min} = 10^8 M_\odot$ scenario, which reaches threshold $J_{\rm LW,21} = 0.1$ at $z = 20.5$. The curves show the power at different redshifts in the same model: $J_{\rm LW,21} \sim 0.001,\,0.01,\,0.1,\,1,$ and $10$ (corresponding to $z = 25.8, \, 23.35, \, 20.5,\, 17.0,$ and $11.05$ from top to bottom). We display the time dependent versions (dashed curves) for the first and last of these. }
\label{delta2lw_fmass}
\end{figure}

Results are depicted in Figures~\ref{delta2lw_z10} and~\ref{delta2lw_fmass}. In the former, we simultaneously vary $M_{\rm min}$ and redshift so that $J_{\rm LW,21} = 0.1$ is fixed, while in the latter we follow a single star formation model over redshift (varying $L_{\rm LW, 21}$).

The normalized scenarios displayed in Figure~\ref{delta2lw_z10} include: $M_{\rm min} = 10^6 M_\odot$, $z=30.15$ (bottom panel) and $M_{\rm min} = 10^8 M_\odot$, $z=20.5$ (top panel). The short-dashed curves represent the light cone versions for both scenarios, using equation~(\ref{timeprofile}) for the flux profile. It is evident that the inclusion of the light cone effect preserves the shape of the power but modestly boosts the amplitude (by a factor of $\sim 2$). We have also separately displayed the 1-halo and 2-halo terms (bottom-most, solid lines) for the $M_{\rm min}=10^6 M_\odot$ scenario so as to gain a sense of when these terms are dominant and to show how they work in tandem to determine the shape of $\Delta^2(k)$. 

It is important to note that the above prescriptions assume that all halos above the mass threshold, $M_{\rm min}$, form stars continuously. Of course, these stars have finite lifetimes, and in the classical Pop III scenario in which each halo undergoes only a short burst of star formation, not all of these stars are going to be `turned on' when we take a snapshot of the fluctuations at a particular point in time. We can account for this simply by incorporating a duty cycle, $f_{\rm duty}$, into our calculation. This addition exclusively affects the one-halo term in equation~\ref{ps_halomodel}; since both $n(m)$ and the effective $f_{\rm coll}$ (which in this model gives the fraction of halos hosting \emph{active} sources) are altered by a factor of $f_{\rm duty}$, the two-halo term remains unchanged.

The two topmost, long-dashed curves in both panels of Figure~\ref{delta2lw_z10} represent the $M_{\rm min} = 10^8 M_\odot$ and $10^6 M_\odot$ scenarios including the light cone effect using two different values of $f_{\rm duty}$. A reasonable estimate for $f_{\rm duty}$ would be the ratio between the average lifetime of a Pop III star, $\tau$, and the Hubble time, $t_H$. The lifetime of these massive stars is believed to be a few million years (Myrs)~\citep{barkloeb2001, brommlarson2004}, though that remains to be directly measured. The topmost curve in both cases assumes an average lifetime of $\tau \sim 3$ Myrs ($f_{\rm duty} = 0.02$ for $M_{\rm min} = 10^6 M_\odot$ and $0.01$ for $M_{\rm min} = 10^8 M_\odot$) while the second curve assumes $\tau \sim 10$ Myrs ($f_{\rm duty} = 0.06$ and $0.04$ for $M_{\rm min} = 10^6 M_\odot$ and $10^8 M_\odot$ respectively). This greatly increases the importance of the one-halo term and so boosts the fluctuations on scales below the LW horizon.  However, note that the mean background intensity also falls by a factor of $f_{\rm duty}$, so these strong fluctuations occur well \emph{before} threshold is reached.

The most striking feature common to all power spectra is the first turnover located at $k_{\rm LW}  \sim 0.06$ cMpc$^{-1}$. This is a strong signature of the LW flux profile, which terminates at $r_{\rm LW} \sim 100$ cMpc ($k_{\rm LW} = 2\pi/r_{\rm LW} \sim 0.06$ cMpc$^{-1}$). Power is smallest for the largest scales and then steadily increases until it reaches $k_{\rm LW}$. In this regime, regions are far outside the LW horizon of each source and so sample independent patches in the radiation field.  The total power is therefore simply proportional to the matter power spectrum multiplied by a mean bias factor (squared).  However, at $k_{\rm LW}$, $\Delta^2$ turns over and begins to fall. This is because such scales sample the variations within $r_{\rm LW}$; if the two points see the same halo populations, their radiation amplitudes will vary together and the fluctuations decrease. On the smallest scales the power turns up and increases monotonically. This indicates where the $P^{\rm 1h} (k)$ term becomes dominant, which occurs on larger scales (smaller $k$) for increasing choice of $M_{\rm min}$ because the sources become more rare.

Note how the signature shifts to slightly higher $k$ in the light cone versions, because the damping scale $r_\star < r_{\rm LW}$. The turnover is also smoothed out as $M_{\rm min}$ increases and the one-halo component begins to dominate at larger scales. This signature is further smoothed by accounting for the duty cycle. The shorter the stellar lifetime (and the smaller the duty cycle), the more amplified the one-halo term will be relative to the unchanged two-halo contribution.

Figure~\ref{delta2lw_fmass} shows the $M_{\rm min} = 10^8 M_\odot$ scenario at several different redshifts.  Curves are labeled according to their normalized $J_{\rm LW,21}$ values. The central curve (blue in the online version) corresponds to the $J_{21} \sim 0.1$ normalized version (z=20.5). Apparent in Figure~\ref{delta2lw_fmass} is the washing out of the turnover at increasing $z$; at high $z$ halos are more rare and the LW background patchier -- consequently the 1-halo term begins to dominate earlier on scales $k < k_{\rm LW}$, thus smoothing out the key signature.

 One important caveat for our model is the assumption of linear bias when computing the 2-halo term in the power spectrum.  For example, consider the $M_{\rm min}=10^8 \ M_\odot$ model, which reaches threshold at $z \sim 20$.  At that time, such a halo has $b \sim 10$.  Thus, even though the rms density fluctuation on $\sim 5$ Mpc scales is $\sim 0.04$, the halo fluctuations are $\sim b \sigma \sim 0.4$, where nonlinear effects are becoming important.  The steep intensity profiles around these sources make clustering somewhat more important for the radiation background, as found by \citet{mesfurl2009}, and probably enhance the fluctuations on moderately small scales by a factor of a few. For example, \citet{mesfurl2009} found from semi-numeric simulations that nonlinear clustering tends to smooth out the signature turnover in the power spectrum of the ionizing background (where it is due to the smaller attenuation length of high-$z$ ionizing photons); see also the discussion in \S \ref{ahncomp} below.
 
 As discussed in \S~\ref{j21}, we assumed $f_\star = 0.1$ here. This is likely to be an upper limit, and it could be much smaller if, for example, the first star to form in each halo suppresses the formation of any others. In this case the radiation field would not reach threshold until \emph{later}, when there are many more halos and hence smaller fluctuations. Our scenarios therefore provide upper limits to the fluctuation amplitude at threshold.
 
 \begin{figure}
 \begin{center}
\resizebox{8cm}{!}{\includegraphics{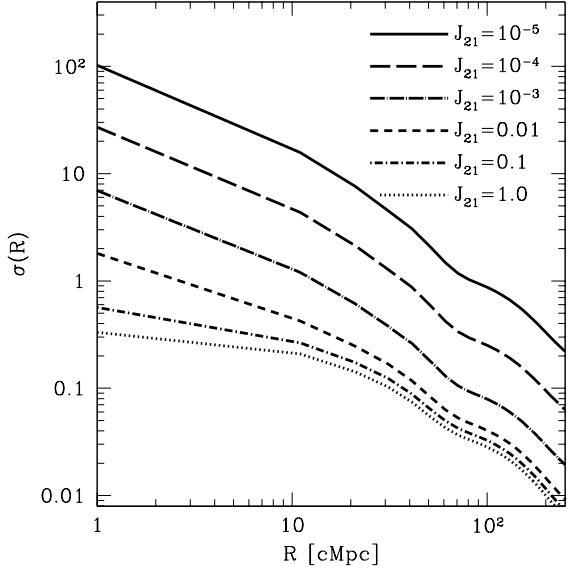}}\\
\end{center}
\caption{The fractional standard deviation, $\sigma(R)$, of the LW background depicted for $M_{\rm min}=10^8 M_\odot$.  From top to bottom: the solid curve is normalized to $10^{-5} J_{21}$ ($z=30$), the long-dashed curve to $10^{-4} J_{21}$ ($z=28$), the dot-long-dashed curve to $10^{-3} J_{21}$ ($z=25.8$), the short-dashed curve to $0.01 J_{21}$ ($z=23.4$), the dot-short-dashed curve to $0.1 J_{21}$ ($z=20.5$), and the dotted curve to $1.0 J_{21}$ ($z=17.0$). We take $f_\star = 0.1$ in these scenarios.}
\label{sigma}
\end{figure}

Given the unusual shapes of these power spectra, we next compute the real-space standard deviation in the intensity to provide better intuition for the amplitude of these fluctuations.
We calculate the fractional standard deviation, $\sigma(R)$, in the following way:
\begin{equation}\label{variance}
\sigma^2(R) = \int dk \frac{k^2}{2\pi^2} P(k) W^{2}_R(k),
\end{equation}
where $W^{2}_R(k)$ is the `window function' or smoothing window over which we consider varying $P(k)$. We employ a simple Gaussian window for computational simplicity:
\begin{equation}
W_R(k) = e^{-k^2R^2/2}.
\end{equation}

We display $\sigma(R)$ in Figure~\ref{sigma} for the choice of $M_{\rm min}=10^8 M_\odot$. From top to bottom, the solid curve is normalized to $10^{-5} J_{21}$ ($z=30.0$), while the others have $J_{\rm LW,21} = 10^{-4},\ 10^{-3}, \ 0.01,\ 0.1,$ and $1$. 

The signature turnover at $k_{\rm LW}$ (see Figures~\ref{delta2lw_z10} and~\ref{delta2lw_fmass}) has been lightly imprinted onto the shape of $\sigma(R)$ in the form of a gentle kink at $R \sim 100$ cMpc. It is evident that at intensity levels nearly approaching, at, and beyond the threshold value, $\sigma(R)$ is small ($\la 1$) down to very small scales ($\sim 1$~cMpc), indicating a fairly uniform background. This suggests that it is unlikely for isolated patches still harboring H$_2$ to exist and foster star-forming minihalos around the threshold. 

\subsection{Fluctuations in the Background at Early Phases}\label{earlyflucs}

Nevertheless, Figure~\ref{sigma} shows that fluctuations in the background are large early on when $J_{\rm LW}$ is well below threshold; $\sigma(8 \ \rm{cMpc}) \sim 20$ for $M_{\rm min}=10^8 M_\odot$ at $z=30$. On the flip side of asking whether or not scattered H$_2$ driven star formation could persist in epochs close to or at threshold, these large fluctuations could indicate that even in epochs for which the background is substantially below threshold there will be patches that are locally at threshold in which H$_2$ cooling is suppressed.

However, in this regime the fluctuations are not gaussian, so the standard deviation $\sigma$ is not a good representation of the importance of the fluctuations.  Moreover, because we primarily care about the radiation intensity at highly clustered sites of other halos -- where star formation is trying to occur -- simply taking a pure spatial average is not necessarily the proper approach (see also \citealt{dijkstra2008}).

\begin{figure}
\begin{center}
\resizebox{8cm}{!}{\includegraphics{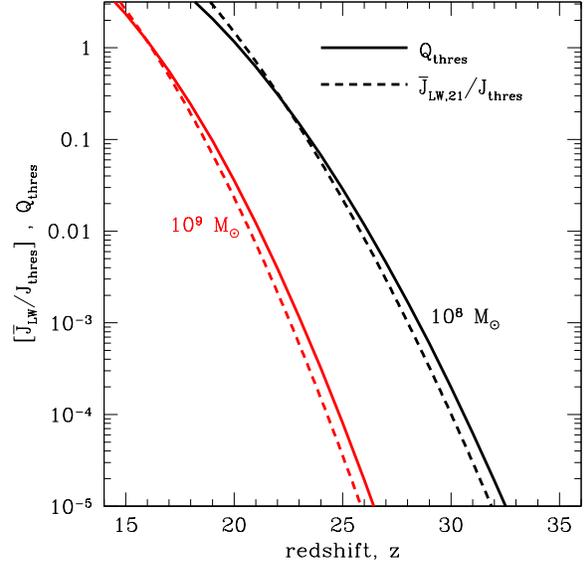}}\\
\end{center}
\caption{The probability, $Q_{\rm thres}$ (solid curves), that a new halo forms within $R_{\rm thres}$ of an existing halo, or within the region irradiated by a local LW intensity above the threshold for H$_2$ suppression. Scenarios include $M_{\rm min} = 10^8 M_\odot$ (rightmost, black curves) and $M_{\rm min} = 10^9 M_\odot$ (leftmost, red curves). Also displayed is the mean LW intensity in units of the threshold level, $\bar J_{\rm LW,21}/J_{\rm thres}$, using $f_\star = 0.1$ and $J_{\rm thres} = 0.1 J_{21}$.}
\label{qprob}
\end{figure}

In order to delve into this new question, we follow the method presented in~\cite{furlloeb2005} with which they calculated the probability that a collapsing halo forms in a region already enriched by galactic winds at high redshift. In contrast, we are interested in calculating the probability that a collapsing halo forms in a region with a LW background above the dissociation threshold. If the sources are very rare, this corresponds to lying within a radius $R_{\rm thres}$ of a LW emitting halo. Within this radius, the `new' halo -- one that has passed the threshold to form stars -- is irradiated by a local LW intensity above the threshold value for suppressing H$_2$ cooling. We can start by calculating the fraction of space contained within $R_{\rm thres}$ of all LW emitting halos, $Q_{\rm thres}'(z)$, assuming that these regions do not overlap:
\begin{equation}
Q_{\rm thres}'(z) = \int_{M_{\rm min}}^{\infty} dm \left( \frac{m}{\bar\rho} \right) \eta(m)n(m),
\end{equation}
where $\eta(m)$ is the ratio of mass irradiated within $R_{\rm thres}$ of a halo with mass $m$ to that halo's mass:
\begin{equation}
\eta(m) = {4 \pi \bar{\rho} R_{\rm thres}^3/3 \over m}.
\end{equation}
For example, for a halo of mass $10^8 M_\odot$ at $z=20$ with $M_{\rm min} = 10^8 M_\odot$ and $f_\star = 0.1$, $R_{\rm thres} \sim 3$ cMpc and $\eta \sim 5.5 \times 10^4$. The corresponding values for $z = 30$ are $R_{\rm thres} \sim 8$ cMpc and $\eta \sim 9.7 \times 10^5$. If the flux profile were a pure $1/r^2$ power law, then $R_{\rm thres} \propto L^{1/2}$, so $\eta \propto m^{1/2}$. In reality, the modulation factor steepens the flux profile, so $\eta$ is closer to flat.

In the limit in which sources are truly isolated, $Q_{\rm thres}'$ would be the total filling factor of threshold regions. However, as more sources appear, their regions will begin to overlap and -- because $\eta$ is an increasing function of $m$ -- grow faster. Because we are only after a crude estimate of this effect, we do not worry about overlap here and use $Q_{\rm thres}'$ as our fiducial estimate. A more sophisticated numerical or Monte Carlo model can easily incorporate this possibility~\citep{dijkstra2008}. The model will, of course, break down when sources become common enough to sit within each other's $R_{\rm thres}$ (indeed, $Q_{\rm thres}'$ is not limited to be less than unity).  Fortunately, in this regime near threshold the halo model approach is perfectly adequate.

Since the newly forming halos are spatially biased and preferentially collapse near existing halos, this expression is not entirely correct. If we consider two halos,  the first a newly formed halo and the second an established LW emitting halo, then the excess probability that the two halos live near each other is quantified by the correlation function, $\xi_{\rm gg}$. To linear order, the correlation function can be written as $\xi_{\rm gg} = b_{\rm new}\bar b_{\rm thres}\xi_{\delta\delta}$, where $b_{\rm new} = b(M_{\rm min})$ and $\bar b_{\rm thres}$ are the biases of the newly collapsed and LW threshold region respectively and $\xi_{\delta\delta}$ is the dark matter correlation function. The mean bias of the LW threshold regions surrounding established halos can be written as:
\begin{equation}
\bar b_{\rm thres} = \frac{\int dm m\eta(m)b(m)n(m)}{\int dm m\eta(m)n(m)}.
\end{equation}

With this in mind, we can approximate the corrected probability that a new halo lives within $R_{\rm thres}$ of an established halo as:
\begin{equation}\label{eqn:q}
Q_{\rm thres} = Q_{\rm thres}' \left[ 1 + b_{\rm new}\bar b_{\rm thres} \xi_{\delta\delta} (R_{\rm thres}) \right].
\end{equation}
We found that these corrections typically boost $Q_{\rm thres}'$ by a factor of $\sim 2$ in our scenarios.

Figure~\ref{qprob} shows $Q_{\rm thres}$ for a choice of $M_{\rm min} = 10^8 M_\odot$ (rightmost, solid black curves) and $M_{\rm min} = 10^8 M_\odot$ (leftmost, solid red curves) with $f_\star = 0.1$. The dashed curves in Figure~\ref{qprob} represent the mean intensity relative to threshold, $\bar J_{\rm LW,21}/J_{\rm thres}$. $Q_{\rm thres}$ increases as redshift decreases and the LW background builds up and becomes more uniform in both scenarios. It is evident from Figure~\ref{qprob} that increasing the choice of $M_{\rm min}$ delays H$_2$ suppression; increasing $M_{\rm min}$ from $10^8 M_\odot$ to $M_{\rm min} = 10^9 M_\odot$ decreases $Q_{\rm thres}$, for example, by a factor of $\sim 33$ at $z \sim 20$. By increasing the mass threshold one eliminates contributions from a host of less massive potential sources, requiring more time for the background to strengthen and boost $Q_{\rm thres}$. Identical calculations using $f_\star = 0.01$ yield values of $Q_{\rm thres}$ that are up to factors of $\sim 15$ smaller.

Also, note that $Q_{\rm thres}$ increases roughly in proportion to $\bar{J}_{\rm LW}$: evidently the mean background provides a good estimate of the volume illuminated by a high intensity of LW radiation. However, note that the quantitative similarity of this filling factor and $\bar{J}_{\rm LW}/J_{\rm thres}$ shown in Fig.~\ref{qprob} is coincidental and does not occur if we, e.g., change our choice of threshold value or $f_\star$.

Once $J_{\rm LW,21}$ reaches threshold, $Q_{\rm thres} \sim 1.2$ (for $M_{\rm min}=10^8 M_\odot$ and $f_\star = 0.1$). At earlier times, the discrete nature of the sources does substantially increase the probability for a new halo to lie within a threshold region over and above what one might naively guess from Figure~\ref{sigma}; for example, when $J_{\rm LW} \sim 0.1 J_{\rm thres}$, $\sim 15\%$ of the halos lie in this regime, even though $\sigma \la 1$ down to very small scales. Nevertheless, we still find that at early times there are relatively few patches above threshold.

\subsection{Comparison to Other Work}\label{ahncomp}

There has been relatively little work on fluctuations in the LW background at high redshifts. The most salient comparison is to \citet{ahn}, who calculated the LW background power spectra using a large-scale radiative transfer simulation with size $R_{\rm box} \sim 50$~cMpc. They resolved halos down to $10^8 M_\odot$ and implemented a simple two-population model for galaxies, in which halos with $M<10^9 M_\odot$ had large ionizing efficiencies and larger halos had more modest efficiencies.  By following the radiative transfer of ionizing photons, they also included the suppression of galaxy formation in halos with $M<10^9 M_\odot$ following reionization, so this higher-efficiency population gradually disappeared.

Although this simulation box should be sufficiently large to include a fair sample of the halo population,\footnote{According to the methods of \citet{barkloeb2004}, the missing large-scale modes only suppress the halo population by a few percent.} it is still much smaller than $r_{\rm LW}$.  \citet{ahn} therefore used a periodic tiling in order to fill out the LW horizon.  Unfortunately, this means that they cannot measure the turnover at $k_{\rm LW}$, nor the regime in which the two-halo component dominates and approaches the straightforward limit $b^2 P^{\rm lin}$, so interpreting their results is somewhat difficult.

We are unable to implement the self-regulated reionization model used by \citet{ahn} in our simpler analytic model, but we have nevertheless made several test calculations to compare our results, choosing a reasonable minimum mass to match theirs at the different redshifts.  Fortunately, \citet{ahn} find that their simulation reaches threshold at $z \approx 16$, when the ionized fraction is less than a percent -- thus we do not expect the self-regulation to be important in the regime of most interest. On the other hand, their box has only one radiation source at $z=19$, so the pre-threshold regime probably suffers from their finite box size even more than expected.

Comparing to their Figure~12, we find less power, by an order of magnitude or so, in the $k \sim 0.3$--$20$~Mpc$^{-1}$ regime probed by their simulations, from $z \sim 16$--$8$.  (They provide only an upper limit at higher redshifts because of the many fewer sources in the box.)  However, our model does help to explain the \emph{shape} of their power spectrum, which (when converted to $\Delta^2$) shows relatively flat power over this range, especially at the lower redshifts.  This is because it lies between the turnover at $k_{\rm LW}$ and the regime in which the one-halo term dominates (see, e.g., the lower-redshift curves in our Fig.~\ref{delta2lw_fmass}).  The flattening becomes more pronounced at lower redshifts as the one-halo term decreases in importance, thanks to the increased source density.

Nonlinear clustering is one likely explanation for the different amplitudes: the simulations can include the fully nonlinear clustering of these sources, while our model ignores them.  \citet{mesfurl2009} did indeed find a boost of power on comparable scales comparing a halo-model implementation of radiation fluctuations to semi-numeric simulations (in this case, the ionizing background at $z \sim 6$).  However, they found a much more modest boost (a factor $\sim 2$), followed by a steepening toward much smaller scales relative to the halo model prediction. Reconciling our results with those of~\cite{ahn} requires a much larger effect. One possible explanation is the amplitude of the source fluctuations, which is $\sim 2$ times larger in the~\citet{ahn} model at $z=20$ than in the~\citet{mesfurl2009} comparison, because the much higher bias of the higher-redshift halos compensates for the smaller fluctuations in the density field.

Another possible explanation is the finite source lifetimes imposed in the numerical simulations, which decreases the number of sources in the box and increases the importance of the one-halo term \citep{ahn,iliev2007}, albeit in a non-uniform manner within the numerical simulation.

In any case, however, both the simulations and analytic models agree that, before the threshold is reached, fluctuations are relatively unimportant. The large-scale uniformity of the background seems robust, in the absence of large-scale modulation to the source population itself (as may be provided by relative velocities between baryons and dark matter; \citealt{tseliakhovich2010}).

In principle, we can also compare our model with the detailed Monte Carlo simulations of~\citet{dijkstra2008}. However, they focus exclusively on times far beyond threshold and very close halo pairs, where our crude approximation no longer applies. 

\section{The Lyman-$\alpha$ Background}\label{lya}

The Lyman-$\alpha$ background imprints fluctuations onto the 21-cm signal~\citep{barkloeb2005} by way of the Wouthuysen-Field effect~\citep{wouthuysen1952, field1958} in which the two hyperfine states of neutral hydrogen are mixed via the absorption and reemission of a Lyman-$\alpha$ photon. Once the first sources in the universe turn on and amalgamate into a Lyman-$\alpha$ background, this effect drives the spin temperature, $T_S$, to the gas temperature, $T_K$, resulting in a nonzero brightness temperature relative to the CMB, $T_b$, and allowing the 21-cm line to become visible. We can write $T_b$ as~\citep{fob}:
\begin{equation}\label{tbright}
T_b(\nu) \approx 9x_{\rm {HI}} (1+\delta)(1+z)^{1/2}\left[ 1-\frac{T_\gamma(z)}{T_S} \right] \left[ \frac{H(z)/(1+z)}{dv_\parallel/dr_\parallel} \right] \ \rm {mK},
\end{equation}
where $x_{\rm HI}$ is the neutral fraction, $(1+\delta)$ is the fractional overdensity of baryons, $T_\gamma(z) = 2.73 (1+z) \ \rm K$ is the brightness temperature of the CMB, and $dv_\parallel/dr_\parallel$ is the gradient of the proper velocity along the line of sight. We can relate $T_\gamma/T_S$ to $T_\gamma/T_K$ with~\citep{fob}:
\begin{equation}\label{tbrightcmb}
\left[ 1-\frac{T_\gamma(z)}{T_S} \right] = \frac{x_c+x_\alpha}{1+x_c+x_\alpha} \left[ 1-\frac{T_\gamma(z)}{T_K} \right],
\end{equation}
where $x_c$ and $x_\alpha$ are the collisional and Lyman-$\alpha$ scattering coupling coefficients.

Observing the 21-cm signal could provide us a window with which to investigate properties of the exotic sources that collectively shaped the nature of this background. This means that the Lyman-$\alpha$ background is a directly observable effect of (nearly) the same photons that make up the LW background. This allows us to measure that feedback process directly rather than having to infer it from modeling star formation in halos, which is quite difficult.

\begin{figure}
\begin{center}
\resizebox{8cm}{!}{\includegraphics{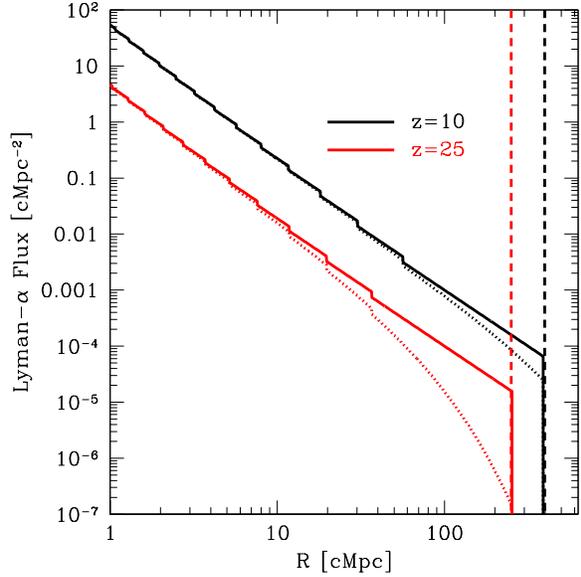}}\\
\end{center}
\caption{The Lyman-$\alpha$ flux profile shown for $z=10$ (black, top curve) and $25$ (red, bottom curve) with $M_{\rm min} = 10^8 M_\odot$. The light cone versions are displayed as dotted curves and the corresponding dashed vertical lines indicate the Lyman-$\alpha$ horizons. The $z=10$ curve is displaced by a factor of $10$ to make it more visible.}
\label{lyaprofile}
\end{figure}

\subsection{The Flux Profile}\label{lyaflux}
Our construction of the Lyman-$\alpha$ flux profile follows~\cite{pritchardfurl}. To calculate the Lyman-$\alpha$ flux originating from a particular source one must consider contributions from all Ly$n$ levels. After absorption (and ignoring recombinations directly to the ground state, which just regenerate the original photons), a fraction, $f_{\rm recycle}$~\citep{hirata2006}, of Ly$n$ photons will be converted into Lyman-$\alpha$ photons via cascades through a series of radiative transitions and will contribute to the total observed Lyman-$\alpha$ flux. The remainder will wind up in the metastable 2$s$ configuration and decay via two-photon emission, resulting in no Lyman-$\alpha$ photon. We use the values of $f_{\rm recycle}$ presented in~\cite{pritchardfurl}. For example, $f_{\rm recycle} = 0.2609, 0.3078$ for $n = 4, 5$ respectively. At large $n$, $f_{\rm recycle}$ asymptotes to a value of $\sim 0.36$. However we note that quantum selection rules ($\Delta L = \pm 1$) forbid a Lyman-$\beta$ photon from producing a Lyman-$\alpha$ photon. 

Since a source can only be separated from the observer by a finite distance before its photons redshift into their nearest Ly$n$ transitions, a photon received at redshift $z$ as a Ly$n$ photon must have been emitted below redshift $z_{\rm max}$:
\begin{equation}
1 + z_{\rm max}(n) = (1 + z) \frac{[1 - (n + 1)^{-2}]}{(1 - n^{-2})}.
\end{equation}

This imprints a set of horizons on the flux profile; horizons become smaller and smaller as you consider higher Ly$n$ levels. The series of horizons results in a step-like structure of the overall profile. We assume the flux from one halo takes the following form:
\begin{equation}\label{fluxlya}
F_{\rm Ly\alpha} \propto \frac{m}{4\pi r^2} \sum_{n}^{n_{\rm max}} \epsilon_{\rm b, \alpha}(\nu_n') f_{\rm recycle}(n)
\end{equation}
where $\epsilon_{\rm b, \alpha}(\nu)$ is the spectral distribution function (defined as the number of photons per baryon emitted at frequency $\nu$ per unit frequency) described by a power law $\epsilon_b(\nu) \propto \nu^{\alpha_s - 1}$. We use the values for $\alpha_s$ presented in~\cite{barkloeb2005} for massive Population III stars and Population II stars. A photon emanating from the source at emission frequency, $\nu'_n$, located at redshift $z'$ is absorbed by level $n$ at redshift $z$:
\begin{equation}
\nu'_n = \nu_n \frac{(1 + z')}{(1 + z)}.
\end{equation}
The sum is ultimately truncated at $n_{\rm max} = 23$ to exclude levels for which the horizon lies within the HII region of a typical (isolated) galaxy~\citep{barkloeb2005a}.

The Lyman-$\alpha$ flux profiles for $z=10$ (top curve) and $25$ (bottom curve) are displayed in Figure~\ref{lyaprofile}. We normalize the curves arbitrarily here in order to focus on the shape as a function of redshift. The Lyman-$\alpha$ horizon distance, $r_{\rm Ly\alpha}$, corresponds to the distance over which a Lyman-$\beta$ photon would redshift into the Lyman-$\alpha$ resonance. This is the maximum range a photon can travel and become a Lyman-$\alpha$ photon; at $z=10$ and $25$, $r_{\rm Ly\alpha}\sim 390$ and $254$ cMpc respectively. In comparison, the LW horizons for those redshifts are $\sim 130$ and $84$ cMpc respectively. We can therefore expect \emph{smaller} fluctuations in the Lyman-$\alpha$ background than in the LW background.  However, note that the difference is not as large as one might otherwise expect because the delay from the light travel time already reduces the importance of distant sources.

The dotted curves represent the light cone corrected profiles; we treat the light cone effect in the same fashion that we amended the LW profile in \S \ref{timedep_lw}. As can be seen in Figure~\ref{lyaprofile}, the light cone curves begin to diverge from the original version on scales $R > 10$ cMpc. The final and largest `step' (and more and more of the smaller steps as you look at higher redshift) in the profile is effectively beveled as the flux begins to prematurely slope downward until it runs into the horizon. This difference morphs the shape of the power spectrum as discussed further in \S \ref{lyaresults}.

\subsection{The Mean 21-cm Background}\label{mean21cm}

We next estimate the redshifts for which the Lyman-$\alpha$ background fluctuations are important when observing the 21-cm signal. Following~\cite{fob} we can write the fractional variation of the brightness temperature of the 21-cm line, $\delta_{21}$, in the following way:
\begin{equation}\label{delta21}
\delta_{21} = \beta\delta_b + \beta_\alpha \delta_\alpha - \delta_{\partial_v},
\end{equation}
where $\delta_b$ is the perturbation in the baryonic density, $\delta_\alpha$ is that for the Lyman-$\alpha$ coupling coefficient  $x_\alpha$, and $\delta_{\partial_v}$ is that for the line of sight peculiar velocity gradient. The expansion coefficients, $\beta_i$, and their evolution over time determine the epochs for which the various perturbations influence the fluctuations in $T_b$. In particular,
\begin{equation}\label{eqn:beta}
\beta = 1 + \frac{x_c}{x_{\rm tot}(1+x_{\rm tot})}
\end{equation}
and
\begin{equation}\label{beta}
\beta_\alpha = \frac{x_\alpha}{x_{\rm tot}(1 + x_{\rm tot})}.
\end{equation}
$\beta_\alpha$ is basically the fractional contribution of the Wouthuysen-Field effect~\citep{wouthuysen1952, field1958} to the coupling, where $x_{\rm tot} \equiv x_c + x_\alpha$ and $x_c$ and $x_\alpha$ are the coupling coefficients for collisions and Lyman-$\alpha$ scattering. For simplicity in our calculations, and for easy comparison to earlier work \citep{barkloeb2005, pritchardfurl}, we ignore all fluctuations except for those due to density ($\beta \delta_b$) and the Lyman-$\alpha$ background ($\beta_\alpha \delta_\alpha$).  We neglect perturbations in the neutral fraction ($\beta_x \delta_x$) and the gas kinetic temperature, $T_K$ ($\beta_T \delta_T$).

The collisional coupling coefficient was calculated as in~\cite{furlanetto2006}. The Lyman-$\alpha$ coupling coefficient can be written as
\begin{equation}\label{xalpha}
x_\alpha = S_\alpha \frac{J_\alpha}{J_\alpha^c},
\end{equation}
where $S_\alpha$ is a correction factor of order unity~\citep{chenmiralda2004, hirata2006, pritchardfurl} that we neglect in our simple model and $J_\alpha$ is the mean Lyman-$\alpha$ intensity.  In a similar fashion to equation~(\ref{jlw}), $J_\alpha$ is given by:
\begin{equation}
J_\alpha(z) = \sum_{n=2}^{n_{\rm max}} \int_z^{z_{\rm max}(n)} dz' f_{\rm recycle}(n) \frac{(1 + z)^2}{4\pi} \frac{c}{H(z')} \epsilon(\nu_n', z').
\end{equation}\label{jla}

The critical intensity, $J_{\alpha,21}^c = 0.66 [(1 + z)/20]$ (in the units of $J_{21}$), corresponds to the threshold level of $J_\alpha$ for which $T_S$ sticks to $T_{K}$~\citep{fob, chenmiralda2004}. How does this threshold intensity compare to the LW intensity at which H$_2$ cooling is suppressed? We have displayed $J_\alpha^c$ in units of $J_{21}$ as the dot-dashed line in Figure~\ref{j21_allz_allm}. The dashed curves in Figure~\ref{j21_allz_allm} represent the calculated average Lyman-$\alpha$ intensities for scenarios with $M_{\rm min} = M_{\rm filter}(z), \ 10^6 M_\odot, \ 10^7 M_\odot,$ and $10^8 M_\odot$ and $f_\star = 0.1$. Notice how these intensities are larger than their LW counterparts (solid curves); the Lyman-$\alpha$ horizon distance is a factor of $\sim 3$ times larger than the LW horizon, which not only allows the Lyman-$\alpha$ background to build up more quickly but also allows for a more uniform background, as discussed in \S~\ref{lyaflux}. 

Around the time that the LW intensity reaches threshold for H$_2$ suppression, $J_\alpha$ is also somewhat higher than that, and hence very close to $J_\alpha^c$. As a result, the 21-cm background is directly sensitive to the physics of cooling; around the time when $T_S$ sticks to $T_K$ numerous minihalos are shutting down stellar production as their H$_2$ supplies are wiped out. Conveniently, this makes the 21-cm background a nearly direct probe of this very interesting epoch in the history of galaxy formation.

\begin{figure}
\begin{center}
\resizebox{8cm}{!}{\includegraphics{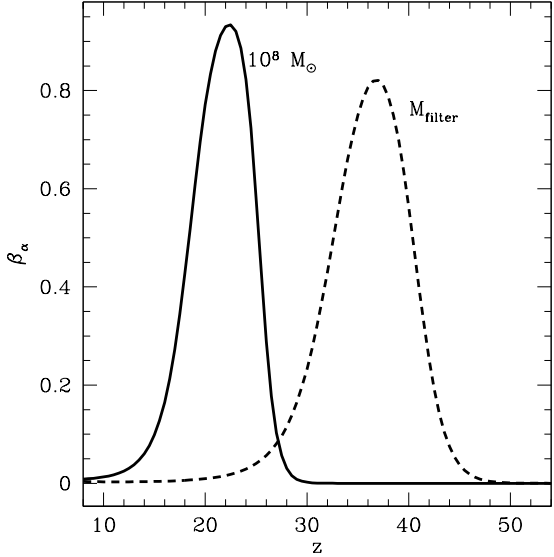}}\\
\end{center}
\caption{The Lyman-$\alpha$ coupling coefficient, $\beta_\alpha(z)$, for $M_{\rm min} = M_{\rm filter}$ (dashed line) and $10^8 M_\odot$ (solid line), assuming $f_\star = 0.1$.}
\label{betaz}
\end{figure}

We present $\beta_\alpha(z)$ in Figure~\ref{betaz} for $M_{\rm min} = 10^8 M_\odot$ and $f_\star = 0.1$. We find that, for $M_{\rm min} = 10^8 M_\odot$, $\beta_\alpha$ peaks at $z \sim 22$ and is significantly nonzero from $z \sim 15 -30$; fluctuations in the Lyman-$\alpha$ background are important over this range. For this scenario, the mean LW background reaches threshold ($J_{\rm LW,21} \sim 0.1$) by $z \sim 21$.

\subsection{Results}\label{lyaresults}

\begin{figure}
\begin{center}
\resizebox{8cm}{!}{\includegraphics{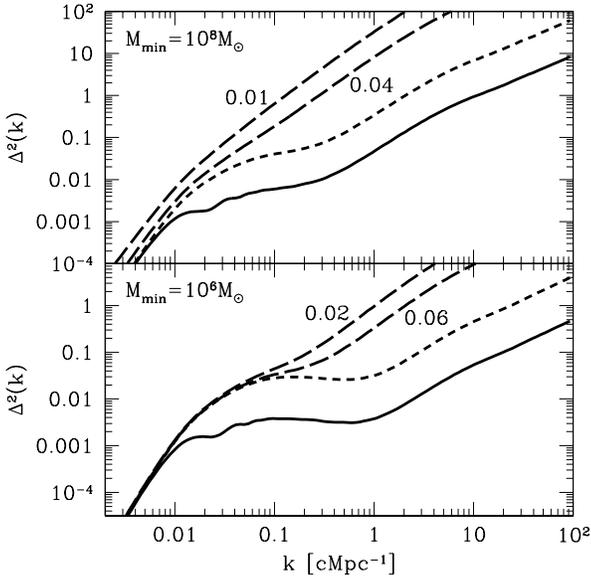}}\\
\end{center}
\caption{Power spectrum of the Lyman-$\alpha$ background, $\Delta^2 (k)$, for scenarios normalized to reach $J_{\rm LW,21} = 0.1$ for ease of comparison with Fig.~\ref{delta2lw_z10}. Scenarios include: $M_{\rm min} = 10^6 M_\odot$ at $z=30.15$ (bottom panel) and $M_{\rm min} = 10^8 M_\odot$ at $z=20.5$ (top panel). The light cone versions for both scenarios are also shown (short-dashed curves) while the original versions are the solid curves. The two, topmost, long-dashed curves include the light cone effect and duty cycle for the $M_{\rm min} = 10^8 M_\odot$ case ($f_{\rm duty} = 0.01$ and $0.04$; labeled $0.01$ and $0.04$ respectively) and the $M_{\rm min} = 10^6 M_\odot$ case ($f_{\rm duty} = 0.02$ and $0.06$).}
\label{delta2lya_threslw}
\end{figure}

Results for the Lyman-$\alpha$ radiation background power spectra are displayed in Figure~\ref{delta2lya_threslw}. For ease in comparison to our LW results, we use the corresponding scenarios from Fig~\ref{delta2lw_z10}, normalized to reach $J_{\rm LW,21} \sim 0.1$. Values for $\beta_\alpha$ in these scenarios are 0.89 for $M_{\rm min} = 10^6 M_\odot$ at $z=30.15$ (bottom panel) and 0.83 for $M_{\rm min} = 10^8 M_\odot$ at $z=20.5$ (top panel). The light cone versions are also shown (short-dashed curves; original versions are the solid curves). The two, long-dashed curves at the top of both panels represent fluctuations for these scenarios including the light cone effect and duty cycle. The topmost curve in both cases assumes an average lifetime of $\tau \sim 3$ Myrs ($f_{\rm duty} = 0.02$ for $M_{\rm min} = 10^6 M_\odot$ and $0.01$ for $M_{\rm min} = 10^8 M_\odot$) while the second curve assumes $\tau \sim 10$ Myrs ($f_{\rm duty} = 0.06$ and $0.04$ for $M_{\rm min} = 10^6 M_\odot$ and $10^8 M_\odot$ respectively).

As discussed in \S~\ref{mean21cm}, the mean Lyman-$\alpha$ intensity, $J_\alpha$, is very nearly the critical intensity, $J_\alpha^c$, around the time that the LW intensity reaches the threshold level for H$_2$ suppression (this can be seen in Figure~\ref{j21_allz_allm}). However, $J_\alpha \propto f_{\rm duty}$: stars are `turned on' for a smaller fraction of the time and thus build the radiation background more slowly. Thus the duty cycle curves in this Figure are \emph{not} at the coupling threshold.  Obviously the fluctuations are boosted on small and mid-range scales (see Figure~\ref{delta2lya_threslw}), but this is not surprising given that we are no longer probing the threshold epochs.

Present in the original models are a series of sequentially damped wiggles, in contrast to the smooth transition of the LW power. These result from the (Fourier transform of the) discontinuous horizon steps present in the Lyman-$\alpha$ flux profile. Unfortunately, these signature wiggles are smoothed out once the light cone effect is applied to the models. As discussed earlier in \S \ref{lyaflux}, the light cone effect bevels out the horizon steps that give the flux profile its distinctive shape, resulting in a more featureless profile and producing a nearly featureless power spectrum.

The power turns over at roughly $k_{\rm Ly\alpha} \sim 2\pi/r_{\rm Ly\alpha}$, where $r_{\rm Ly\alpha}$ is the Lyman-$\alpha$ horizon distance (discussed above in \S \ref{lyaflux}), except for the light cone versions whose turnovers shift to somewhat higher $k$. In addition, the amplitude of the light cone Lyman-$\alpha$ power is roughly a factor of $2$ smaller than the corresponding amplitudes for the LW background for scales of $k \sim 0.1$ cMpc$^{-1}$. This is likely a symptom of the larger Lyman-$\alpha$ horizon distances, which are $\sim 3 r_{\rm LW}$ for these redshifts. Furthermore, the light cone effect on the Lyman-$\alpha$ power is stronger than the corresponding effects on the LW fluctuations. The Lyman-$\alpha$ models corrected for halo growth over time are boosted in amplitude by a factor of $\sim 7$, while the LW models receive a boost by a factor of $\sim 1.5$. The large Lyman-$\alpha$ horizon allows points to `see' more halos, bolstering the light cone effect on the most distant sources.

\subsection{The 21-cm Signal}

Finally, armed with the fluctuations in the radiation background and the fluctuations in the baryon density (computed with the linear power spectrum on these scales), we can estimate the 21-cm signal itself. Referring back to equation~(\ref{delta21}), we consider fluctuations in $T_b$ sourced by perturbations in the matter density, Wouthuysen-Field coupling (or the Lyman-$\alpha$ flux), and radial velocity gradient of the gas. All of these fluctuations are isotropic except for the velocity fluctuation, which introduces an anisotropy to the power spectrum and can be written as $\delta_{\partial_v}(k) = -\mu^2\delta_b$~\citep{bharadwajali2004}, where $\mu$ is the cosine of the angle between the wavenumber $\mathbf{k}$ of the Fourier mode and the line of sight. This enables the total power spectrum for $T_b$ to be separated into powers of $\mu^2$~\citep{barkloeb2005a}:
\begin{equation}
P_{T_b}(\mathbf{k}) = \mu^4P_{\mu^4}(\mathbf{k}) + \mu^2P_{\mu^2}(\mathbf{k}) + P_{\mu^0}(\mathbf{k}).
\end{equation}

The $\mu^2$ term, which can be written as~\citep{barkloeb2005a}
\begin{equation}
P_{\mu^2}(k) = 2\mu^2[ \beta P_\delta(k) + \beta_\alpha P_{\delta-\alpha}(k)],
\end{equation}
contains contributions of density-induced fluctuations in the Lyman-$\alpha$ flux, where $P_{\delta-\alpha}(k)$ is the cross-power spectrum for the matter density and Lyman-$\alpha$ radiation background. In our halo model, this is very easy to calculate (c.f.,~\citealt{cooraysheth}):
\begin{eqnarray}
P_{\delta-\alpha}(k) & = & P_{\delta-\alpha}^{1h}(k) + P_{\delta-\alpha}^{2h}(k), \mbox{where} \\
P_{\delta-\alpha}^{1h}(k) & = & \int_{M_{\rm min}}^{\infty} dm n(m) \left(\frac{m}{\bar \rho f_{\rm coll}} \right)\left(\frac{m}{\bar \rho}\right) |u_\delta(k|m)| |u_{\alpha}(k|m)| \\
P_{\delta-\alpha}^{2h}(k) & = & \left[ \int_{0}^{\infty} dm n(m) \left(\frac{m}{\bar \rho} \right) u_\delta(k|m) b(m) \right] \\
& & {} \times \left[ \int_{M_{\rm min}}^{\infty} dmn(m) \left(\frac{m}{\bar \rho f_{\rm coll}} \right) u_{\alpha}(k|m) b(m) \right] P^{\rm lin}(k),
\label{crossps}
\end{eqnarray}
where $u_\delta$ is the halo density profile that describes the distribution of mass within each halo.

\begin{figure}
\begin{center}
\resizebox{8cm}{!}{\includegraphics{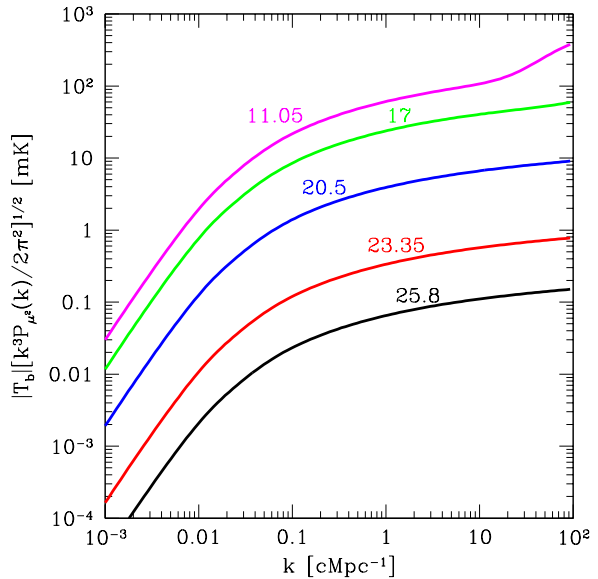}}\\
\end{center}
\caption{Power spectrum of $T_b$ for the 21-cm transition, $|T_b|[k^3P_{\mu^2}(k)/2\pi^2]^{1/2}$ (in mK), for scenarios with $M_{\rm min} = 10^8 M_\odot$ for ease of comparison with Fig.~\ref{delta2lw_fmass}. Scenarios include: $z=11.05, 17.0, 20.5, 23.35,$ and $25.8$ from top to bottom. }
\label{ptb}
\end{figure}

Therefore, $P_{\mu^2}(k)$ can easily be used to investigate fluctuations in the Lyman-$\alpha$ background at high redshift. We display $P_{\mu^2}(k)$ in Figure~\ref{ptb} for $M_{\rm min} = 10^8 M_\odot$ at redshifts $z = 11.05, 17.0, 20.5, 23.35,$ and $25.8$ (from top to bottom) so as to correspond to the scenarios presented in Figure~\ref{delta2lw_fmass}. We find that fluctuations in $T_b$ increase with decreasing redshift and decrease with scale. The increase in amplitude levels off once $J_\alpha$ reaches the critical intensity, $J_\alpha^c$ (which, for the scenario in Figure~\ref{ptb}, occurs around $z \sim 18$). At this point the Lyman-$\alpha$ coupling is saturated ($x_{\rm tot} \gg 1$) and the 21 cm fluctuations become insensitive to the fluctuations in the Lyman-$\alpha$ background. One can also pick out the 'one-halo' term kicking in on small scales for these later epochs (the $z=11.05$ curve in Figure~\ref{ptb}).

The amplitude and overall shape of our power spectra are comparable to those presented in~\cite{pritchardfurl}, who used the same model to describe the Lyman-$\alpha$ flux but calculated $P_{\mu^2}$ using a linear transfer function. The amplitudes from our model also agree with those from~\cite{barkloeb2005}, although the shape of the power differs because they neglected the effects of atomic cascading (by assuming $f_{\rm recycle} = 1$) in their calculations. 

Unfortunately, these power spectra do not display any sort of tell-tale signature feature such as the distinct LW turnover as can be seen in Figures~\ref{delta2lw_z10} and~\ref{delta2lw_fmass} or the wiggles from the horizon steps, because the density-induced fluctuations wash them out. This agrees with previous work; recently,~\cite{vonlanthen2011} showed with a radiative transfer numerical simulation that the horizon steps present in the Lyman-$\alpha$ flux profile left imprints in the differential brightness temperature profile just after the first luminous sources turned on. However, as time progressed and more new sources began contributing to the Lyman-$\alpha$ background, the steps were effectively wiped out. 

\section{Summary \& Conclusions}\label{summary}

In this paper, we have modeled fluctuations in the LW and Lyman-$\alpha$ radiation backgrounds using a variation of the halo model. First, we calculated the LW power spectrum and found that the power is characterized by an abrupt cut-off at the LW horizon distance, $r_{\rm LW}$; the power turns over at the horizon wavenumber, $k_{\rm LW}$, unless the sources are so rare that the one-halo term dominates (i.e., correlations are determined by the flux profiles of individual sources). 

We found that the fluctuations in the background are weak and should not lead to substantial spatial fluctuations in H$_2$ content. Once a population of low-mass halos produces enough stars to generate a threshold LW background large enough to destroy their own H$_2$ reservoirs used for cooling, star formation can only proceed in larger halos with more substantial reservoirs. Our model predicts that, by the time this threshold is reached, fluctuations in the intensity field will be quite small, so this transition will be rather homogeneous across the entire Universe.  

Though we found fluctuations in the background to be small around threshold, on the flip side we also found them to be large in those early epochs during which the background was approaching threshold. This could indicate the presence of patches of collapsing halos that are locally above threshold and H$_2$ suppressed. Taking into account the bias of sources, we crudely approximated the probability that a new halo lives in such a patch and found that even though source clustering substantially increases this probability, it is still relatively small. Thus, at this time, the majority of newly-forming halos can continue to cool via H$_2$, even in the presence of established galaxies~\citep{dijkstra2008}.

Eventually the first stars will build up the LW background by enough to suppress their own star formation; the process will terminate when only those halos above the atomic cooling threshold can form stars.  However, the mode of star formation in these halos is very different from the minihalos that produce the first stars, so the transition to higher-mass halos has important consequences for the global star formation history \citep{ohhaiman2002}. This transition from Population III to Population II star formation is of course extremely complex, and we have examined only a small part of it.  The formation of very massive Population III stars requires two physical conditions: (1) metal-free gas and (2) a reservoir of H$_2$ that allows the gas to cool \citep{brommlarson2004}. We have not examined the first condition, but the slow speed of galactic winds (compared to the speed of light) guarantees that metal enrichment will be very inhomogeneous, and pristine pockets of gas could persist until very low redshifts \citep{scann2002, furlloeb2005}. In contrast, the LW background is spatially uniform and so will induce a rapid, homogeneous transformation in the fundamental processes of star formation, even when metal enrichment remains inhomogeneous. This will induce a shift from very massive Population III.1 stars to less massive -- but still primordial -- Population III.2 stars that require atomic cooling.

Next we considered the fluctuations in the Lyman-$\alpha$ background in a similar fashion. The Lyman-$\alpha$ flux profile imprints a series of wiggles on the shape of the power, corresponding to the series of horizon steps that characterize the profile. Unfortunately, unlike the LW case, these signature wiggles are washed out once we account for halo growth over time. We found that the amplitude of the Lyman-$\alpha$ power is smaller than that for the LW background by a factor of $\sim 2$ for scales larger than $\sim 15$ cMpc. The smaller fluctuations are due to the large Lyman-$\alpha$ horizon distance, $r_{\rm Ly\alpha} \sim 3 r_{\rm LW}$, allowing the halos to `see' further.

We used our model for the fluctuations in the Lyman-$\alpha$ background to generate power spectra for the brightness temperature, $T_b$, of the 21-cm signal. We find our values to be in good agreement with previous estimates~\citep{pritchardfurl, barkloeb2005} that used a very different approach to estimate the radiation field fluctuations. We do not see a signature feature present in the power in contrast to the distinct LW turnover.

Our relatively simple model, though convenient and efficient, has a number of caveats that compromise accuracy. Most notably, we neglect nonlinear effects on the backgrounds by relying on a linear approximation for the halo-halo correlation function. Thus, our models are only good down to the scales for which this linear approximation still holds. In addition, we assumed a uniform star forming efficiency, $f_\star$, for all minihalos in our calculations. As the halo population grows in size and complexity in the later epochs, variations in galactic properties -- sourced partly by the suppression of H$_2$ cooling, but also due to a myriad of other factors -- will complicate our simple treatment. After comparing our results to the previous simulation from~\cite{ahn} we find that the shapes can be well matched even with the limited dynamic range of the Ahn values (the signature turnover is not covered in their range). Although our amplitudes disagree by a factor of $\sim 10$ at the lower redshifts, we both find gentle fluctuations in the background around the time it reaches threshold, implying that `safe' patches sheltering isolated sources of H$_2$ are rare by this epoch.

Furthermore, we have neglected the effects of X-rays in our models. If the first luminous sources had hard spectra that extended out to X-rays, these X-rays would have far reaching effects due to their large mean free paths -- they could, by catalyzing the formation of new H$_2$, potentially counteract H$_2$ photodissociation by the growing UV background \citep{mcdowell61, hrl96, haiman00}. This could stall the transition of Population III to Population II stars by allowing minihalos to continue forming new stars, altering the make-up of the sources responsible for reionization. 

Finally,~\citet{tseliakhovich2010} argue that a new nonlinear effect must be considered in structure formation: the supersonic relative velocity between dark matter and baryons can suppress the matter power spectrum near the baryonic Jeans scale, altering the abundance and clustering properties of the first dark matter halos. This effect could be accounted for in a future version of our simple model by introducing a modulation factor to the halo mass function. In fact,~\citet{dalalpen10} argue that the resulting fluctuations in the radiation background could strongly affect the 21-cm absorption power spectrum. However, they assume that the velocity exerts a very strong effect on galaxy formation, which may be at odds with more detailed numerical simulations~\citep{maio2010, stacy2010}. In any case, such a large scale modulation will add more power to the radiation field and may have important implications for the homogeneity of the LW and Lyman-$\alpha$ backgrounds, since these acoustic features appear on comparable scales to the LW horizon ($\sim 100$~Mpc).
\\

This research was partially supported by the David and Lucile Packard Foundation, the Alfred P. Sloan Foundation, and by NASA through the LUNAR program. The LUNAR consortium (http://lunar.colorado.edu), headquartered at the University of Colorado, is funded by the NASA Lunar Science Institute (via Cooperative Agreement NNA09DB30A) to investigate concepts for astrophysical observatories on the Moon.  

%
%
\bibliography{bibfile1}

\end{document}